\documentclass[epj]{svjour2}
\usepackage{graphics}
\usepackage{color}

\usepackage{graphicx}
\usepackage{epsf} 
\usepackage{slashed}
\usepackage{amsmath}
\usepackage{amssymb}

\def\be{\begin{equation}}
\def\ee{\end{equation}}
\def\ba{\begin{eqnarray}}
\def\ea{\end{eqnarray}}

\def\sfrac#1#2{{\textstyle \frac{#1}{#2}}}

\begin{document}
\title{New low-$Q^2$ measurements of the  $\gamma^\ast N \to \Delta(1232)$ 
Coulomb quadrupole form factor, pion cloud parametrizations
 and Siegert's theorem}
\author{G.~Ramalho  %\inst{1} 
%\thanks{\emph{email:} gilberto.ramalho2013@gmail.com}%
}                     % Do not remove
\institute{Laborat\'orio de F\'{i}sica Te\'orica e Computacional -- LFTC, \\
Universidade Cruzeiro do Sul, 01506-000, S\~ao Paulo, SP, Brazil}
\date{Received: date / Revised version: date}
\abstract{
The novel measurements of the  $\gamma^\ast N \to \Delta(1232)$ 
Coulomb quadrupole form factor in the range $Q^2=0.04$--0.13 GeV$^2$ 
changed the trend of the previous data. 
With the new data the electric and Coulomb form factors are both
in remarkable agreement with  estimates of the pion cloud contributions to the 
quadrupole form factors at low $Q^2$.  
The pion cloud contributions to the electric and Coulomb 
form factors can be parametrized 
by the relations $G_E \propto G_{En}/\left(1 + \frac{Q^2}{2M_\Delta(M_\Delta-M)}\right)$
and  $G_C \propto G_{En}$, where $G_{En}$ is the neutron electric form factor, and 
$M$, $M_\Delta$ are the nucleon and $\Delta$ masses, respectively.
Those parametrizations are in 
full  agreement with Siegert's theorem, 
which states that $G_E= \sfrac{M_\Delta-M}{2M_\Delta} G_C$
at the pseudothreshold, when $Q^2=-(M_\Delta -M)^2$,
and improve previous parametrizations.
Also a small valence quark component 
estimated  by a covariant quark model contributes to this agreement.
The combination of the new data with the new parametrization for $G_E$ concludes 
an intense period of studying the $\gamma^\ast N \to \Delta(1232)$ quadrupole form factors at low $Q^2$,
with the agreement between theory and data.
   % end of PACS codes
} %end of abstract
\maketitle

\section{Introduction}

The first excited state of the nucleon, the
$\Delta(1232)$, is an exceptional system
in the context of the strong interactions (QCD).
It  dominates the electro-excitations of the nucleon
and nucleon pion-production reactions~\cite{NSTAR,Aznauryan12b,Pascalutsa07b}.
The study of the $\Delta(1232)$ internal structure 
ruled by the quark-gluon degrees of freedom and quark-antiquark excitations, 
interpreted as meson cloud, 
reveals that the  $\gamma^\ast N \to \Delta(1232)$ transition 
is predominantly a magnetic transition~\cite{NSTAR,Aznauryan12b,Bernstein03}.
The magnetic dipole form factor $G_M$
is dominated  by valence quark effects, particularly at
large momentum transfer squared, $Q^2$,
and can be explained even by quark models 
based on a symmetric structure for the nucleon and 
the $\Delta(1232)$~\cite{JDiaz07,Capstick90,NDelta,Lattice,NDeltaD,Eichmann12,Segovia13}.

The   $\gamma^\ast N \to \Delta(1232)$ transition is also 
characterized by two 
sub-leading quadrupole form factors:
the electric ($G_E$) and the Coulomb ($G_C$) 
form factors~\cite{NSTAR,Aznauryan12b,Pascalutsa07b,NDeltaD,LatticeD,Jones73}.
The  non-zero values for the quadrupole form factors 
are interpreted as a consequence of the deviation of the $\Delta(1232)$
from a spherical 
shape~\cite{Pascalutsa07b,Bernstein03,Capstick90,Glashow79,Krivoruchenko91,Isgur82,Deformation,Quadrupole-1,Quadrupole-2,Becchi65,Stave08,Buchmann00b}.
Contrary to the case of the magnetic form factor,
estimates of the electric and the Coulomb quadrupole form factors
based on quark models predict only a small 
fraction of the values measured~\cite{NSTAR,Pascalutsa07b,JDiaz07,Capstick90,NDeltaD,Buchmann00b,Tiator04}.
There are, however, evidence that the missing strength of 
the quadrupole form factors
in quark models is due to meson cloud or quark-antiquark 
effects~\cite{LatticeD,Tiator04,SatoLee,QpionCloud-1,QpionCloud-2,QpionCloud-3,Kamalov01}.
   
Estimates based on the limit of a large number of colors ($N_c$)
and $SU(6)$ quark models with symmetry breaking, indicate in fact,
that in the low-$Q^2$ region the $\gamma^\ast N \to \Delta(1232)$ 
quadrupole form factors are dominated by 
pion cloud effects~\cite{Pascalutsa07b,Pascalutsa07a,Buchmann97a,Grabmayr01,Buchmann04,Buchmann02,Buchmann09a}.
Parametrization of pion cloud contributions 
based on large $N_c$ have been proposed to explain 
the empirical data~\cite{Pascalutsa07a,Buchmann09a,Siegert-ND}.
Those parametrization are very close to the 
empirical data~\cite{Pascalutsa07a,Blomberg16a}.
In some cases, small valence quark contributions 
help to improve the description 
of the data~\cite{JDiaz07,LatticeD,Kamalov01,Siegert-ND}.

There are, however, some limitations 
associated with those parametrizations.
Until the last few years
the parame-trization for $G_C$ was in disagreement with the low-$Q^2$ data.
In addition, there is a conflict 
between the pion cloud parametrizations 
and Siegert's theorem~\cite{Pascalutsa07a,Siegert-ND},
which relates the quadrupole form factors $G_E$ and $G_C$
at the pseudothreshold, when $Q^2=-(M_\Delta-M)^2$.

Recently, new data for the Coulomb quadrupole form factor 
became available in the region $Q^2=0.04$--0.13 GeV$^2$ 
from the experiments at Jefferson Lab/Hall A~\cite{Blomberg16a}.
The new data compare extraordinarily well 
with an improved large $N_c$ estimate of the pion cloud 
contributions to the quadrupole form factors,
discussed in the present work, 
when valence quark contributions estimated by  
a covariant quark model are also included.
The new parame-trizations for the quadrupole form factors 
satisfy Siegert's theorem~\cite{Jones73,Siegert-ND,Tiator-1,Tiator-2,Drechsel2007}.

Siegert's theorem states that 
when the $\Delta(1232)$ and the nucleon 
are both at rest, one has~\cite{Jones73,Siegert-ND,SiegertD,Siegert} 
\ba
G_E (Q_{pt}^2)= \kappa G_C(Q^2_{pt}),
\label{eqSiegert1}
\ea
where $\kappa = \sfrac{M_\Delta -M}{2 M_\Delta}$ and 
$Q^2_{pt}=-(M_\Delta-M)^2$.
The condition $Q^2= Q^2_{pt}$ defines the pseudothreshold, 
when the photon three-momentum ${\bf q}$
vanishes ($|{\bf q}|=0$).

\begin{figure}[t]
\vspace{.6cm}
\centerline{\mbox{
\includegraphics[width=3.1in]{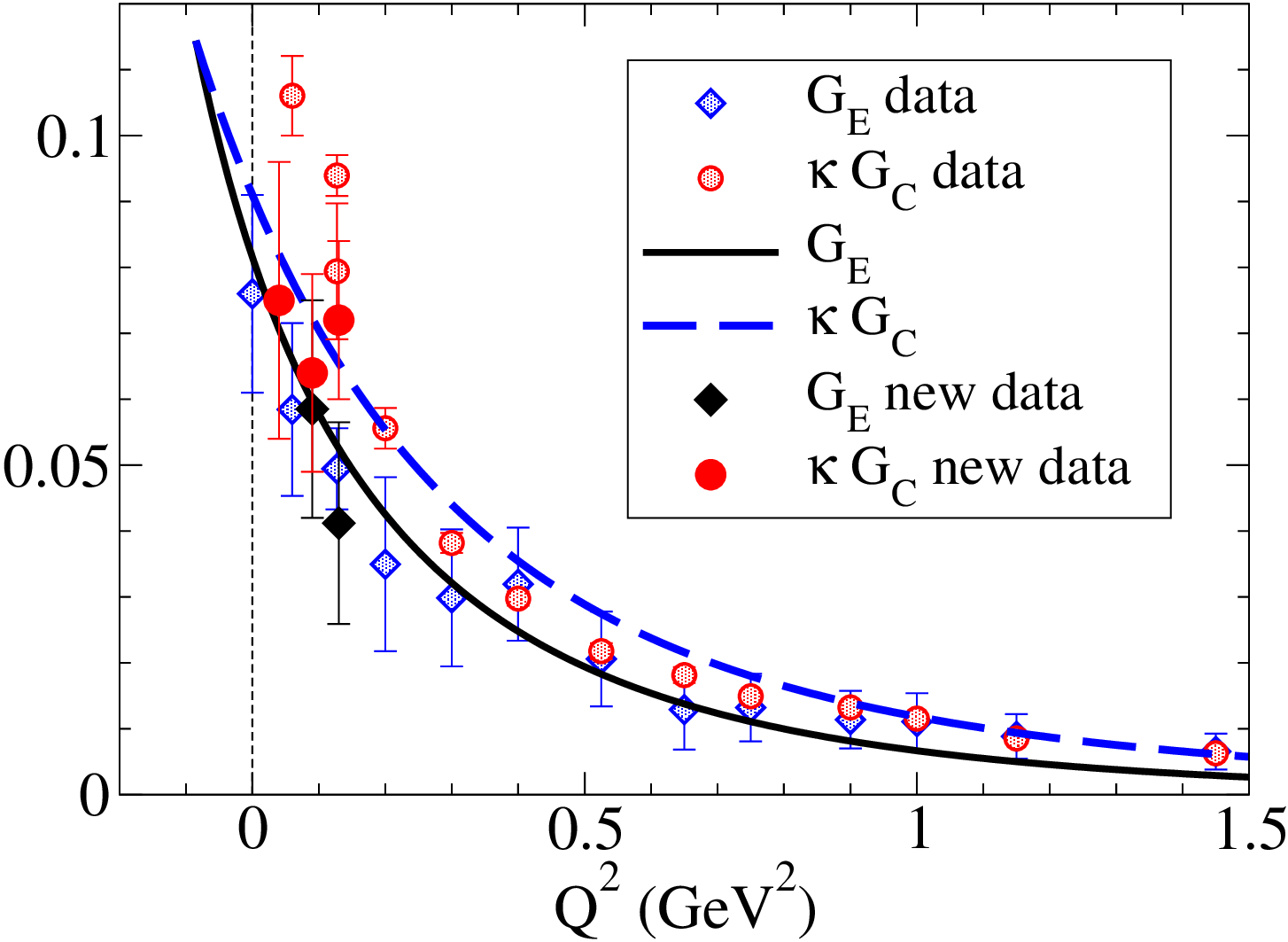}}}
\caption{\footnotesize
$G_E$ and $G_C$ form factors.
Data from refs.~\cite{Stave08,Blomberg16a,Sparveris13,MIT_data,Jlab_data1,Jlab_data2,PDG}.
Recall that $\kappa = \sfrac{M_\Delta-M}{2M_\Delta}$.
}
%\vspace{-1cm}
\label{figTotal}
\end{figure}

The remarkable agreement between the  
parametrizations for the quadrupole form factors ($G_E$ and $G_C$)
and the data  can be observed in fig.~\ref{figTotal}.
The data are from refs.~\cite{Stave08,Blomberg16a,Sparveris13,MIT_data,Jlab_data1,Jlab_data2,PDG}.
Note in particular the excellent 
agreement with the new data from JLab/Hall A~\cite{Blomberg16a}
(solid circles and diamonds). 
The results for $G_C$ are multiplied by 
$\kappa$ for convenience.
In the figure one can notice the convergence 
of the two lines at the lowest $Q^2$ point (pseudothreshold) 
proving the consistency with Siegert's theorem.
The pion cloud parametrizations
for the form factors $G_E$ and $G_C$ discussed next,
as well as  the valence quark contributions discussed later,
contribute to this success.
The inclusion of the valence quark contributions 
compensates the underestimation associated with  
the pion cloud parametrizations~\cite{Pascalutsa07a,Siegert-ND,Blomberg16a,SiegertD}.

\section{Framework}

The internal structure of the baryons can be interpreted 
using as a combination of the large $N_c$ limit, 
with $SU(6)$ quark models 
with two-body exchange currents~\cite{Buchmann04,Buchmann09a}.
The $SU(6)$ symmetry breaking induces an asymmetric distribution 
of charge in the nucleon which generates non-zero results 
for the neutron electric form factor
as shown in constituent quark models 
such as the Isgur-Karl model~\cite{Isgur82,IsgurRefs-1,IsgurRefs-2,IsgurRefs-3} 
and others~\cite{Pascalutsa07b,Buchmann97a,Grabmayr01}.
Using the $SU(6)$ symmetry  breaking 
one can show that the 
$\gamma^\ast N \to \Delta(1232)$
quadrupole moments are proportional 
to the neutron square charge radius 
($r_n^2$)~\cite{Krivoruchenko91,Pascalutsa07a,Buchmann97a,Grabmayr01,Buchmann02,Dillon99,Buchmann02b}.

Using the low $Q^2$ expansion of the 
neutron electric form factor, $G_{En} \simeq -\sfrac{1}{6} r_n^2 Q^2$, 
we can represent the $Q^2$ dependence 
of the quadrupole form factors 
in the form~\cite{Pascalutsa07a,Buchmann97a,Grabmayr01,Buchmann04,Buchmann02,Buchmann09a}:
\ba
& &
G_E^\pi (Q^2) = \left(\frac{M}{M_\Delta} \right)^{3/2} 
\frac{M_\Delta^2 -M^2}{2 \sqrt{2}} 
\frac{\tilde G_{En}(Q^2)}{ 1 + \frac{Q^2}{2 M_\Delta (M_\Delta-M)}}, \nonumber \\
& & \label{eqGE1} \\
& &
G_C^\pi (Q^2) = \left(\frac{M}{M_\Delta} \right)^{1/2} 
\sqrt{2} M_\Delta M \tilde G_{En}(Q^2),
\label{eqGC1}
\ea
where $\tilde G_{En}= G_{En}/Q^2$.

The interpretation of the previous relations as 
the pion cloud contributions is the consequence 
the relations between quadrupole form factors 
and $r_n^2$ based on constituent quark models with 
two-body exchange currents.
The effects of those currents can be interpreted 
as pion/meson contributions~\cite{Grabmayr01,Buchmann04,Buchmann02,Buchmann09a}.
The interpretation is also valid in large $N_c$ limit, 
where the form factors $G_E$ and $G_C$ appear as 
higher orders in $1/N_c^2$, comparared to $G_M$~\cite{Pascalutsa07a,Jenkins02}.

Equations (\ref{eqGE1})-(\ref{eqGC1}) were derived directly from 
the large $N_c$ limit~\cite{Pascalutsa07a},
apart from the denominator of the factor $\tilde G_{En}$ in eq.~(\ref{eqGE1}).
This denominator is included in the present work 
in order to satisfy Siegert's theorem~(\ref{eqSiegert1}), exactly.
Note that in the limit $Q^2 \to 0$ the extra factor 
reduces to the unit, and we recover the original result 
from large $N_c$ limit~\cite{Pascalutsa07a}.
At the pseudothreshold: 
$1 + \sfrac{Q^2}{2 M_\Delta (M_\Delta-M)} = \sfrac{M_\Delta + M}{2M_\Delta}$, 
which leads directly to eq.~(\ref{eqSiegert1}).
Since in the large $N_c$ limit 
$M_\Delta -M = {\cal O}(1/N_c)$, and $M_\Delta = {\cal O}(N_c)$, 
the present form for $G_E^\pi$ corresponds to a correction ${\cal O}(1/N_c^2)$
relative to the original form of $G_E^\pi$ 
presented in ref.~\cite{Pascalutsa07a}, 
at the pseudothreshold.

In a previous work~\cite{Siegert-ND}, a similar expression 
was considered for $G_E^\pi$, which describes Siegert's theorem 
with an error of the order $1/N_c^4$.
The new expression for $G_E^\pi$ improves the previous result
with the exact description of Siegert's theorem
(all orders of $1/N_c$).
Compared to the form presented in ref.~\cite{Siegert-ND},
we include a correction ${\cal O}(1/N_c^4)$
at the pseudothreshold\footnote{Since we can write: 
$1 + \sfrac{Q_{pt}^2}{2M_\Delta(M_\Delta -M)} =
1 + \sfrac{Q_{pt}^2}{M_\Delta^2 -M^2} + {\cal O}\left(\frac{1}{N_c^4} \right)$,
the present form for $G_E^\pi$ differs from the result form 
ref.~\cite{Siegert-ND} by a term 
${\cal O}(1/N_c^4)$.
}.

Works based on quark models
have shown that Siegert's theorem 
can be violated when the current operators 
associated with the charge density and current densities 
are truncated in different 
orders~\cite{Capstick90,Drechsel84,Weyrauch86,Bourdeau87,Buchmann98}.
The main conclusion of those works is that a consistent 
calculation with current conservation requires 
the inclusion of processes beyond the impulse approximation 
at the quark level (one-body-currents)~\cite{Buchmann98}.
It is then necessary to include
higher-order terms, such as two-body currents,
in order to satisfy Siegert's theorem~\cite{Buchmann98}.
As mentioned above, those currents are 
associated with the pion cloud contributions.

To describe the neutron electric form factor we consider 
the Galster parametrization~\cite{Galster71}:
\ba
G_{En}(Q^2) = - \mu_n \frac{a \tau_N}{1 + d \tau_N} G_D,
\label{eqGEn}
\ea
where $\mu_n= -1.913$ is the neutron magnetic moment,
$\tau_N= \sfrac{Q^2}{4 M^2}$, $G_D= 1/(1+ Q^2/0.71)^2$ 
is the dipole form factor, and $a$, $d$ are two dimensionless parameters.
In fig.~\ref{figTotal}, we use $a=0.9$ and $d=2.8$, 
a parametrization that describes very well the neutron 
electric form factor data. 
We assume that eq.~(\ref{eqGEn}) holds for $Q^2 <0$,
because we expect $G_{En}$ to be described by a smooth function 
near $Q^2=0$, and also because the range 
of extrapolation is small, since $Q_{pt}^2 \simeq -0.1$ GeV$^2$.
Similar extrapolations are considered in refs.~\cite{Tiator-1,Tiator-2,Drechsel2007}.

We consider the  Galster parametrization (\ref{eqGEn})
because of its simplicity and also because of the 
limited precision of the  $G_E$ and $G_C$ below 0.3 GeV$^2$.
Other phenomenological parametrizations with a similar number of parameters 
can also be considered~\cite{Kelly02,Friedrich03,Kaskulov04,Bertozzi72,Platchkov90,Gentile11}.
More sophisticated parametrizations, 
with a larger number of parameters have been derived 
based on dispersion relations and  
chiral perturbation theory~\cite{Hammer04,Belushkin07,Lorenz12}.
In a separated work we study a new class of parametrizations 
for $G_{En}$~\cite{InPreparation}.

Some care should be taken with the use of 
the relations  (\ref{eqGE1})-(\ref{eqGC1}),
since in principle they should not be interpreted exclusively as pion cloud 
contributions, because in the empirical parametrization 
of $G_{En}$ one includes all possible contributions,  
including also contributions due to valence quark effects.
We note, however, that in an exact $SU(6)$ model 
the contribution from the valence quarks associated 
with one-body currents vanishes~\cite{Buchmann09a,Dillon99,Lichtenberg78,Close79}.
Thus, in an approximated $SU(6)$ symmetry, 
one can still expect that the quark-antiquark contributions 
are the dominant effect for $G_{En}$ and $r_n^2$~\cite{Buchmann09a,Grabmayr01}.
In those circumstances, one can use eqs.~(\ref{eqGE1})-(\ref{eqGC1})
to estimate the pion cloud contributions to the quadrupole form factors.
Examples of models with pion cloud/sea quark dominance can be found in 
refs.~\cite{Buchmann00b,Buchmann97a,Buchmann91a,Christov96,Lu98}.

\section{Results}

The theoretical estimates presented in fig.~\ref{figTotal} 
are compared with data from Mainz~\cite{Stave08,Sparveris13},
MIT-Bates~\cite{MIT_data} and Jefferson Lab~\cite{Jlab_data1,Jlab_data2} 
for finite $Q^2$,
and the world average from the 
Particle Data Group at $Q^2=0$~\cite{PDG} (empty diamonds and circles). 
The new data at $Q^2=0.06$, 0.13 GeV$^2$ for $G_E$
and $Q^2=0.04$, 0.06, 0.13 GeV$^2$ for $G_C$
are from JLab/Hall A~\cite{Blomberg16a} 
(solid diamonds and circles).
To convert the 
new data for the electromagnetic ratios $R_{EM}\equiv - \sfrac{G_E}{G_M}$
and $R_{SM}\equiv - \sfrac{|{\bf q}|}{2M_\Delta}\frac{G_C}{G_M}$ 
into $G_E$ and $G_C$, we use the MAID2007 parametrization for 
$G_M$: $G_M=3 \sqrt{1 + \tau}  (1+ a_1 Q^2)  e^{-a_4 Q^2} G_D$,
where $\tau= \sfrac{Q^2}{(M_\Delta + M)^2}$, $a_1= 0.01$ GeV$^{-2}$
and $a_4= 0.23$ GeV$^{-2}$~\cite{Drechsel2007}.
The larger error bars associated with the new data 
are mainly the consequence of the different model descriptions 
of the background~\cite{Blomberg16a}.

The pion cloud contributions for 
the  $\gamma^\ast N \to \Delta(1232)$ quadrupole form factors 
given by eqs.~(\ref{eqGE1})-(\ref{eqGC1}) 
can be complemented by small 
valence quark contributions to the respective form factors 
(around 10\%, near $Q^2=0$).
As discussed in ref.~\cite{Siegert-ND}, those contributions 
are naturally consistent with Siegert's theorem.
The valence quark contributions 
to the $\gamma^\ast N \to \Delta(1232)$
quadrupole form factors are produced by
the high angular momentum components 
in the nucleon and/or $\Delta(1232)$ wave functions.
As a consequence of the orthogonality between 
the nucleon and  $\Delta(1232)$ states, 
the valence quark contributions to the quadrupole 
form factors vanish at the pseudothreshold
and the Siegert's theorem condition 
is trivially satisfied~\cite{NDeltaD,LatticeD,Siegert-ND}.
The validity of Siegert's theorem depends 
then only on the pion cloud contribution.
It is for that reason that the parametrizations 
(\ref{eqGE1})-(\ref{eqGC1}) are particularly useful.

\begin{figure}[t]
\vspace{.6cm}
\centerline{\mbox{
\includegraphics[width=3.1in]{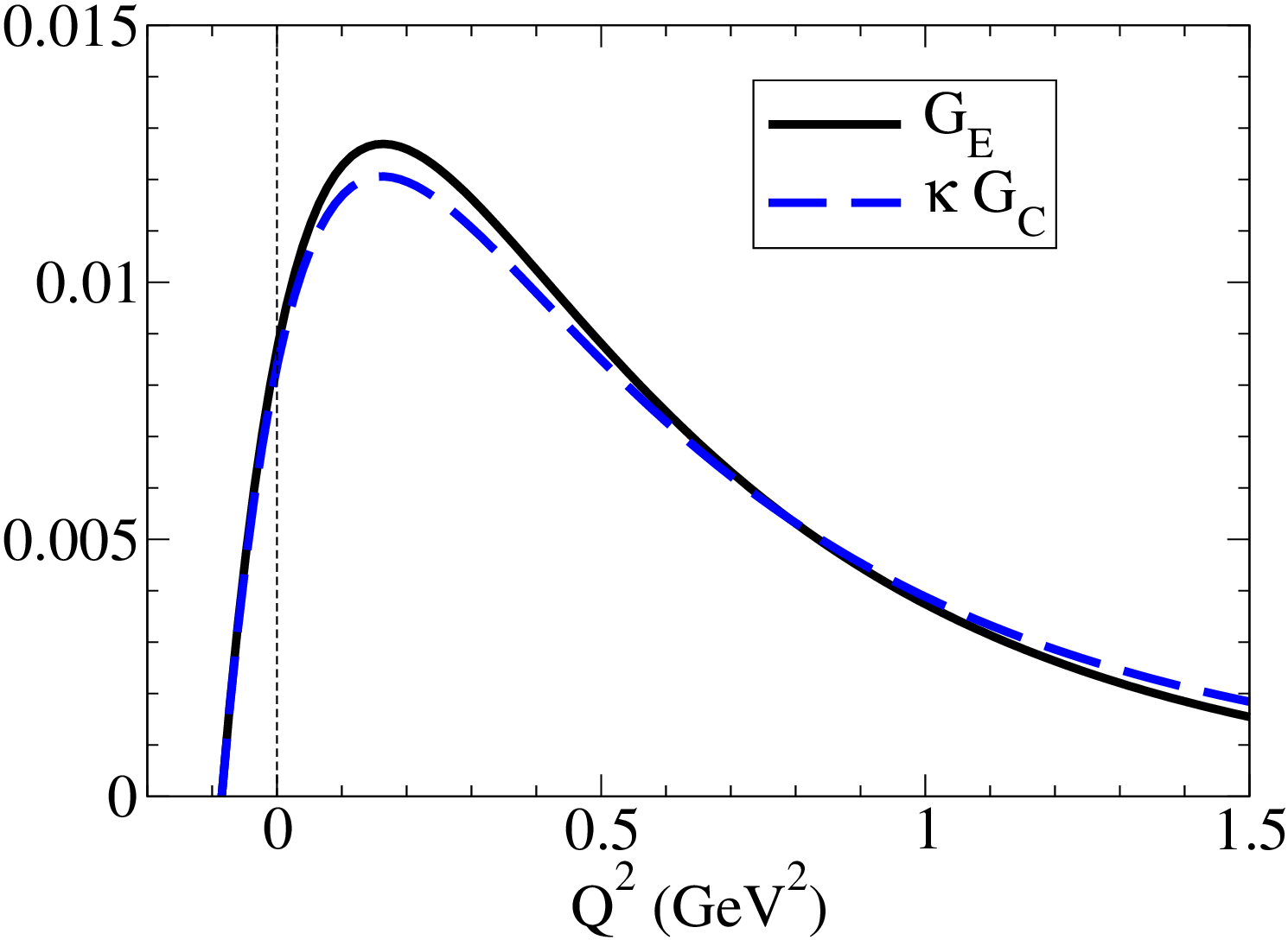}}}
\caption{\footnotesize
Valence quark contributions for $G_E$ and $G_C$ 
according to ref.~\cite{LatticeD}.
Recall that $\kappa = \sfrac{M_\Delta-M}{2M_\Delta}$.}
%\vspace{-1cm}
\label{figValence}
\end{figure}

To estimate the valence quark contribution we use the covariant spectator 
quark model~\cite{NDelta,Lattice,NDeltaD,LatticeD,Nucleon,Omega},
because it is covariant and it is also consistent with 
lattice QCD simulations~\cite{Alexandrou08}.
In ref.~\cite{LatticeD}  the valence quark contributions  
are calculated using a extrapolation from 
lattice QCD results with large pion masses,
to the physical point~\cite{Lattice,LatticeD,Omega}.
The contributions to the quadrupole form 
factors are the consequence of quark $D$-wave
components on the $\Delta(1232)$ wave function~\cite{NDeltaD,LatticeD}.
The free parameters of the model associated with 
the two possible $D$-states are the mixture coefficients 
and parameters associated with the shape of the radial wave functions.
All the free parameters of the model 
are fixed by the lattice data~\cite{LatticeD}.
In the lattice QCD simulations with large pion masses the
meson cloud effects are very small, and the physics associated
with the valence quarks can be better calibrated~\cite{Lattice,LatticeD,Omega}.
The results of the valence quark contributions 
to the form factors $G_E$ and $G_C$ estimated by the model 
from ref.~\cite{LatticeD} 
are presented in fig.~\ref{figValence}.
The results displayed correspond to a 0.72\% mixture 
for both $D$-states.
Note, in the figure, that the quadrupoles form factors follow 
the approximated relation $G_E \simeq \kappa G_C$.

The present estimates of the valence quark contributions 
to the quadrupole form factors can be compared 
with other estimates from the bare core contributions,
such as the Sato-Lee~\cite{JDiaz07,SatoLee} and 
DMT models~\cite{Kamalov01}.
It is important to note, however, that
those parametrizations are not consistent with the constraints 
from quark models, since the quadrupole 
form factors must vanish at the pseudothreshold, 
as a consequence of the orthogonality between the states~\cite{Siegert-ND}.

% FIG. 3

The final results for the 
electric and Coulomb quadru-pole form factors,
presented in fig.~\ref{figTotal} 
are then the sum of the pion cloud parametrizations (\ref{eqGE1})-(\ref{eqGC1}) 
and the valence quark contributions from the covariant 
spectator quark model~\cite{LatticeD}.
It is worth mentioning that the
combination of the two effects is fundamental to 
the agreement between theory and data.
This happens because the modified form of $G_E^\pi$
decreases the estimate of $G_E$, which is compensated by 
the inclusion of the valence quark component.
Also for $G_C$, the addition of the valence quark 
component is essential to reproduce the magnitude 
of $G_C$ below 0.5 GeV$^2$~\cite{SiegertD}.

In fig.~\ref{figTotal}, one can also notice that there are some discrepancy 
between the new JLab/Hall A data 
and the previous measurements from MAMI and 
MIT-Bates~\cite{Stave08,Sparveris13,MIT_data}
below $0.15$ GeV$^2$.
The magnitude of the form factor $G_C$ is smaller than 
in previous measurements.
The result at $Q^2=0.06$ GeV$^2$ from MAMI~\cite{Stave08} 
is inconsistent with the new results at $Q^2=0.04$ and 0.09 GeV$^2$.
This discrepancy has been identified as a consequence of the 
procedure used to calculate the resonant amplitudes
from the cross sections~\cite{Blomberg16a}.

In the present work, we restrict our study to the low $Q^2$ region, 
because we cannot expect the pion cloud parametri-zations (\ref{eqGE1})-(\ref{eqGC1}) 
to be valid for arbitrary large values of $Q^2$,
since they are derived from the low-$Q^2$ relation $G_{En} \simeq - \sfrac{1}{6} r_n^2 Q^2$.
One can then assume that for large values of $Q^2$,
eqs.~(\ref{eqGE1})-(\ref{eqGC1})
are modified according to $G_E^\pi \to G_E^\pi/(1+ Q^2/\Lambda_E^2)^2$
and $G_C^\pi \to G_C^\pi/(1+ Q^2/\Lambda_C^2)^4$, 
where $\Lambda_E$ and $\Lambda_C$ are large momentum cutoff parameters.
In those conditions, the form factors $G_E$ 
and $G_C$ would be, at large $Q^2$, dominated by the 
valence quark contributions, as predicted by perturbative QCD, 
with falloffs: 
$G_E \propto 1/Q^4$ and $G_C \propto 1/Q^6$~\cite{Carlson-1,Carlson-2}.

\begin{figure}[t]
\vspace{.6cm}
\centerline{\mbox{
\includegraphics[width=3.1in]{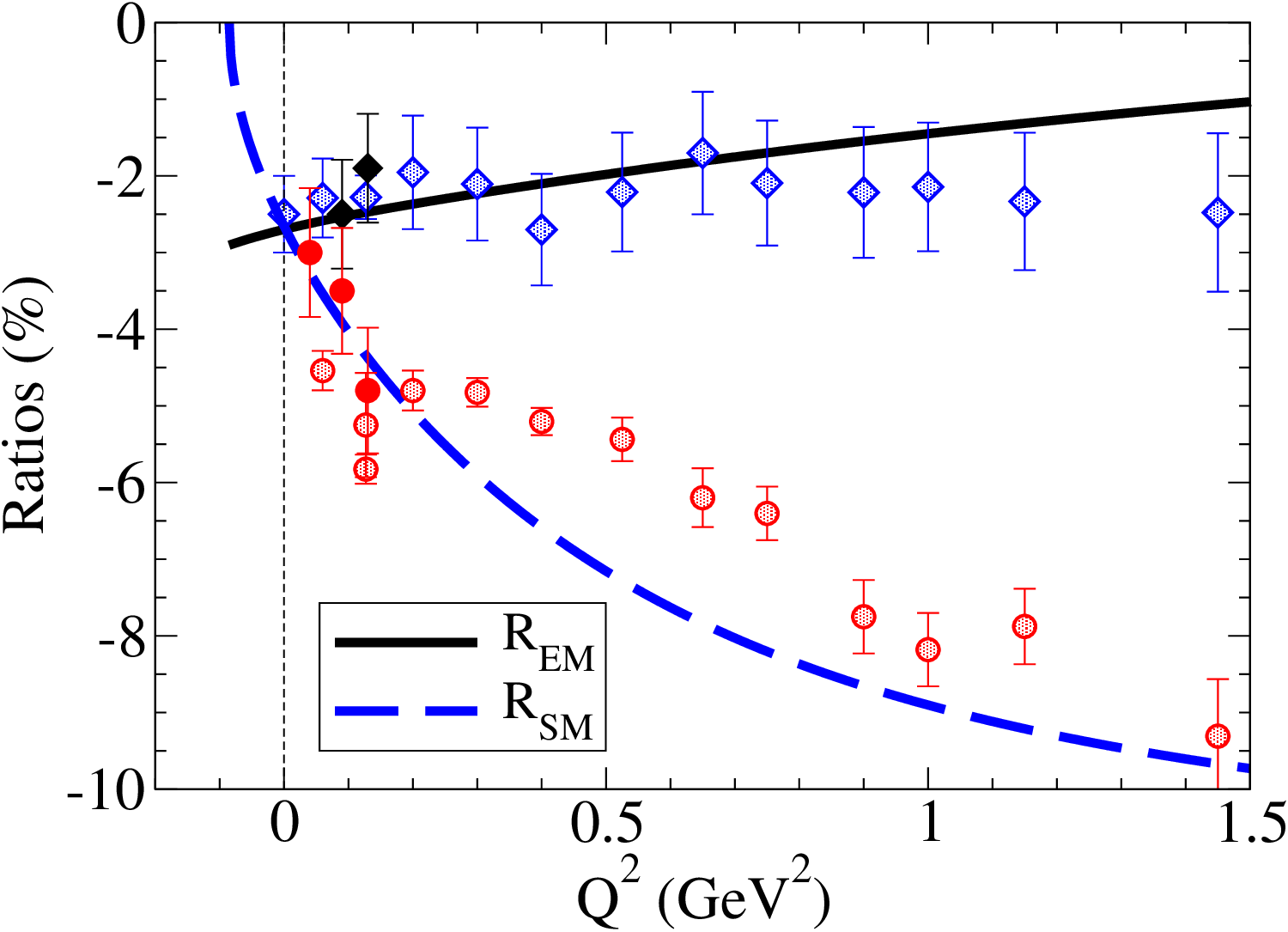}}}
\caption{\footnotesize
Ratios $R_{EM}$ and $R_{SM}$.
Data from refs.~\cite{Stave08,Blomberg16a,Sparveris13,MIT_data,Jlab_data1,Jlab_data2,PDG}.}
%\vspace{-1cm}
\label{figREM-RSM0}
\end{figure}

Using the previous results for $G_E$ and $G_C$  
we can also calculate the electromagnetic ratios $R_{EM}$ and $R_{SM}$,
and compare the results with the measured data.
To estimate $G_M$ we use the MAID2007 parametrization~\cite{Drechsel2007}.
The comparisons are presented in fig.~\ref{figREM-RSM0}, up to 1.5 GeV$^2$.
In that region one can observe some deviation between the model   
and the $R_{SM}$ data for $Q^2=0.3$--0.8 GeV$^2$.
This result seems to indicate that the 
pion cloud parametrization for $G_C$ may not be very accurate 
as the parametrizations 
presented in other works~\cite{Pascalutsa07a,Buchmann09a}.
It is worth mentioning, however, that those works 
use parametrizations of $G_M$ based on 
relations with the nucleon form factors, 
and not the empirical parametrizations of $G_M$, as in the present study.

Another important point to discuss 
is the value of the functions $R_{EM}$ and $R_{SM}$ when $Q^2=0$. 
In fig.~\ref{figREM-RSM0} the two functions are very close at $Q^2=0$. 
The numerical result is $R_{SM} -R_{EM} \simeq 0.05$\%.
This result corroborates the large $N_c$ estimate:
$R_{EM}(0) = R_{SM}(0)$, apart from terms ${\cal O}(1/N_c^2)$~\cite{Pascalutsa07a}.
In our framework, the previous result 
is the consequence of  the combination between 
the relation for the pion cloud parametrizations 
at $Q^2=0$: 
$G_E^\pi = \frac{M_\Delta^2-M^2}{4M_\Delta} G_C^\pi$, 
and the approximated relation 
between the valence quark contributions: 
$G_E \simeq \frac{M_\Delta-M}{2M_\Delta} G_C$  
$\approx \frac{M_\Delta^2-M^2}{4M_\Delta^2} G_C$~\footnote{
 One can write 
$\sfrac{M_\Delta-M}{2M_\Delta} = \sfrac{2 M_\Delta}{M_\Delta + M}  
\sfrac{M_\Delta^2-M^2}{4M_\Delta^2}$, 
obtaining 
$\sfrac{M_\Delta-M}{2M_\Delta} \simeq \frac{M_\Delta^2-M^2}{4M_\Delta^2}$, 
apart from relative corrections of the order 
$\sfrac{M_\Delta -M}{M_\Delta + M}= {\cal O}\left(\sfrac{1}{N_c^2}\right)$.},
as observed in fig.~\ref{figValence}.

The consistence of the new data can be tested in the near future 
by lattice QCD simulations with pion masses near 
the physical point~\cite{Alexandrou11}.
Meanwhile, for simulations not too far from the pion physical mass, 
one can test the compatibility between 
lattice QCD simulations and empirical data
using chiral effective field theories
and chiral quark models~\cite{Pascalutsa06,Gail06,Agadjanov14,Faessler06}.

\section{Outlook}

To summarize, in this work we present parametrizations 
for the pion cloud contributions to the $\gamma^\ast N \to \Delta(1232)$ 
quadrupole form factors 
that are fully consistent with Sie-gert's theorem.
When we combine those parametrizations with 
a consistent calculation of the valence quark contributions, 
we obtain 
an excellent description of the available data, including in particular, 
the most recent measurements of $G_E$ and $G_C$ at low $Q^2$.
Since the valence quark components are extrapolated from lattice QCD,
and the pion cloud parametrizations are determined by $G_{En}$,
our final results are genuine predictions.

The understanding of the proprieties 
of the quadrupole form factors at low $Q^2$
has been a challenge, 
since the derivation of the relations between 
$\gamma^\ast N \to \Delta(1232)$ qua-drupoles 
and  the neutron square charge radius
in the context of constituent quark models,
$SU(6)$ symmetry and large $N_c$ limit,
and since the first measurements of the quadrupole ratios 
in the modern accelerators~\cite{firstMeas-1,firstMeas-2,firstMeas-3}.
Combining two features, namely, the new data for the qua-drupole form factors 
and a new parametrization for the pion cloud contribution for $G_E$,
we have achieved at last a consistent description  
of the  $\gamma^\ast N \to \Delta(1232)$ quadrupole form factors at low $Q^2$.

\section*{Acknowledgments}
%\begin{acknowledgments}
The author thanks Nikos Sparveris  for useful discussions
and Kazuo Tsushima for comments and suggestions.
This work was supported by the Funda\c{c}\~ao de Amparo \`a 
Pesquisa do Estado de S\~ao Paulo (FAPESP):
project no.~2017/02684-5, grant no.~2017/17020-BCO-JP. 
%\end{acknowledgments}

%\input{bibloND3X}


\begin{thebibliography}{00}


\bibitem{NSTAR} 
  I.~G.~Aznauryan {\it et al.},
  %``Studies of Nucleon Resonance Structure in Exclusive Meson Electroproduction,''
  Int.\ J.\ Mod.\ Phys.\ E {\bf 22}, 1330015 (2013).
  %% doi:10.1142/S0218301313300154
  %%[arXiv:1212.4891 [nucl-th]].
  %%CITATION = doi:10.1142/S0218301313300154;%%
  %63 citations counted in INSPIRE as of 18 Dec 2015

\bibitem{Aznauryan12b} 
  I.~G.~Aznauryan and V.~D.~Burkert,
  %``Electroexcitation of nucleon resonances,''
  Prog.\ Part.\ Nucl.\ Phys.\  {\bf 67}, 1 (2012).
  %%doi:10.1016/j.ppnp.2011.08.001
  %%[arXiv:1109.1720 [hep-ph]].
  %%CITATION = doi:10.1016/j.ppnp.2011.08.001;%%
  %113 citations counted in INSPIRE as of 28 Aug 2017


\bibitem{Pascalutsa07b} 
  V.~Pascalutsa, M.~Vanderhaeghen and S.~N.~Yang,
  %``Electromagnetic excitation of the Delta(1232)-resonance,''
  Phys.\ Rept.\  {\bf 437}, 125 (2007).
  %% doi:10.1016/j.physrep.2006.09.006
  %%[hep-ph/0609004].
  %%CITATION = doi:10.1016/j.physrep.2006.09.006;%%
  %199 citations counted in INSPIRE as of 10 May 2016


\bibitem{Bernstein03} 
  A.~M.~Bernstein,
  %``Deviation of the nucleon shape from spherical symmetry: Experimental status,''
  Eur.\ Phys.\ J.\ A {\bf 17}, 349 (2003).
  %%doi:10.1140/epja/i2002-10176-7
  %%[hep-ex/0212032].
  %%CITATION = doi:10.1140/epja/i2002-10176-7;%%
  %35 citations counted in INSPIRE as of 28 Aug 2017




\bibitem{JDiaz07} 
  B.~Julia-Diaz, T.-S.~H.~Lee, T.~Sato and L.~C.~Smith,
  %``Extraction and Interpretation of gamma N ---> Delta Form Factors within a Dynamical Model,''
  Phys.\ Rev.\ C {\bf 75}, 015205 (2007).
  %% doi:10.1103/PhysRevC.75.015205
  %%[nucl-th/0611033].
  %%CITATION = doi:10.1103/PhysRevC.75.015205;%%
  %58 citations counted in INSPIRE as of 10 May 2016





\bibitem{Capstick90} 
  S.~Capstick and G.~Karl,
  %``$\frac{E_{1+}}{M_{1+}}$ and $\frac{S_{1}}{M_{1+}}$ and Their $Q^2$ Dependence in $\gamma_\nu N \to \Delta$ With Relativized Quark Model Wave Functions,''
  Phys.\ Rev.\ D {\bf 41}, 2767 (1990).
  %% doi:10.1103/PhysRevD.41.2767
  %%CITATION = doi:10.1103/PhysRevD.41.2767;%%
  %71 citations counted in INSPIRE as of 01 Jan 2016



\bibitem{NDelta} 
  G.~Ramalho, M.~T.~Pe\~na and F.~Gross,
  %``A Covariant model for the nucleon and the Delta,''
  Eur.\ Phys.\ J.\ A {\bf 36}, 329 (2008).
  %% doi:10.1140/epja/i2008-10599-0
  %%[arXiv:0803.3034 [hep-ph]].
  %%CITATION = doi:10.1140/epja/i2008-10599-0;%%
  %44 citations counted in INSPIRE as of 25 Dec 2015





\bibitem{Lattice} 
  G.~Ramalho and M.~T.~Pe\~na,
  %``Nucleon and gamma N ---> Delta lattice form factors in a constituent quark model,''
  J.\ Phys.\ G {\bf 36}, 115011 (2009).
  %% doi:10.1088/0954-3899/36/11/115011
  %%[arXiv:0812.0187 [hep-ph]].
  %%CITATION = doi:10.1088/0954-3899/36/11/115011;%%
  %31 citations counted in INSPIRE as of 11 May 2016



\bibitem{NDeltaD} 
  G.~Ramalho, M.~T.~Pe\~na and F.~Gross,
  %``D-state effects in the electromagnetic N Delta transition,''
  Phys.\ Rev.\ D {\bf 78}, 114017 (2008).
  %% doi:10.1103/PhysRevD.78.114017
  %%[arXiv:0810.4126 [hep-ph]].
  %%CITATION = doi:10.1103/PhysRevD.78.114017;%%
  %44 citations counted in INSPIRE as of 25 Dec 2015





\bibitem{Eichmann12} 
  G.~Eichmann and D.~Nicmorus,
  %``Nucleon to Delta electromagnetic transition in the Dyson-Schwinger approach,''
  Phys.\ Rev.\ D {\bf 85}, 093004 (2012).
  %%doi:10.1103/PhysRevD.85.093004
  %%[arXiv:1112.2232 [hep-ph]].
  %%CITATION = doi:10.1103/PhysRevD.85.093004;%%
  %36 citations counted in INSPIRE as of 19 Sep 2017

\bibitem{Segovia13} 
  J.~Segovia, C.~Chen, C.~D.~Roberts and S.~Wan,
  %``Insights into the gamma* N -> Delta transition,''
  Phys.\ Rev.\ C {\bf 88}, 032201 (2013).
  %% doi:10.1103/PhysRevC.88.032201
  %%[arXiv:1305.0292 [nucl-th]].
  %%CITATION = doi:10.1103/PhysRevC.88.032201;%%
  %24 citations counted in INSPIRE as of 19 Sep 2017


\bibitem{LatticeD} 
  G.~Ramalho and M.~T.~Pe\~na,
  %``Valence quark contribution for the gamma N ---> Delta quadrupole transition extracted from lattice QCD,''
  Phys.\ Rev.\ D {\bf 80}, 013008 (2009).
  %% doi:10.1103/PhysRevD.80.013008
  %%[arXiv:0901.4310 [hep-ph]].
  %%CITATION = doi:10.1103/PhysRevD.80.013008;%%
  %32 citations counted in INSPIRE as of 25 Dec 2015





\bibitem{Jones73} 
  H.~F.~Jones and M.~D.~Scadron,
  %``Multipole gamma N Delta form-factors and resonant photoproduction and electroproduction,''
  Annals Phys.\  {\bf 81}, 1 (1973).
  %% doi:10.1016/0003-4916(73)90476-4
  %%CITATION = doi:10.1016/0003-4916(73)90476-4;%%
  %233 citations counted in INSPIRE as of 17 Dec 2015





\bibitem{Glashow79} 
  S.~L.~Glashow,
  %``The Unmillisonant quark,''
  Physica A {\bf 96}, 27 (1979).
  %%CITATION = PHYSA,A96,27;%%
  %85 citations counted in INSPIRE as of 02 Sep 2017




\bibitem{Krivoruchenko91} 
  M.~I.~Krivoruchenko and M.~M.~Giannini,
  %``Quadrupole moments of the decuplet baryons,''
  Phys.\ Rev.\ D {\bf 43}, 3763 (1991).
  %% doi:10.1103/PhysRevD.43.3763
  %%CITATION = doi:10.1103/PhysRevD.43.3763;%%
  %26 citations counted in INSPIRE as of 13 May 2017





\bibitem{Isgur82} 
  N.~Isgur, G.~Karl and R.~Koniuk,
  %``D Waves in the Nucleon: A Test of Color Magnetism,''
  Phys.\ Rev.\ D {\bf 25}, 2394 (1982).
  %% doi:10.1103/PhysRevD.25.2394
  %%CITATION = doi:10.1103/PhysRevD.25.2394;%%
  %187 citations counted in INSPIRE as of 05 Sep 2017








\bibitem{Deformation} 
  G.~Ramalho, M.~T.~Pe\~na and A.~Stadler,
  %``The shape of the $\Delta$ baryon in a covariant spectator quark model,''
  Phys.\ Rev.\ D {\bf 86}, 093022 (2012).
  %% doi:10.1103/PhysRevD.86.093022
  %%[arXiv:1207.4392 [nucl-th]].
  %%CITATION = doi:10.1103/PhysRevD.86.093022;%%
  %11 citations counted in INSPIRE as of 17 May 2016


\bibitem{Quadrupole-1} 
  G.~Ramalho, M.~T.~Pe\~na and F.~Gross,
  %``Electric quadrupole and magnetic octupole moments of the Delta,''
  Phys.\ Lett.\ B {\bf 678}, 355 (2009).
  %% doi:10.1016/j.physletb.2009.06.052
  %%[arXiv:0902.4212 [hep-ph]].
  %%CITATION = doi:10.1016/j.physletb.2009.06.052;%%
  %26 citations counted in INSPIRE as of 31 May 2016
  
\bibitem{Quadrupole-2} 
  G.~Ramalho, M.~T.~Pe\~na and F.~Gross,
  %``Electromagnetic form factors of the Delta with D-waves,''
  Phys.\ Rev.\ D {\bf 81}, 113011 (2010).
  %% doi:10.1103/PhysRevD.81.113011
  %%[arXiv:1002.4170 [hep-ph]].
  %%CITATION = doi:10.1103/PhysRevD.81.113011;%%
  %37 citations counted in INSPIRE as of 12 Aug 2016




% Small E2, C2


\bibitem{Becchi65} 
  C.~Becchi and G.~Morpurgo,
  %``Vanishing of the E2 part of the N$^∗_{33}$ → N + γ amplitude in the non-relativistic quark model of “elementary” particles,''
  Phys.\ Lett.\  {\bf 17}, 352 (1965).
  %% doi:10.1016/0031-9163(65)90563-9
  %%CITATION = doi:10.1016/0031-9163(65)90563-9;%%
  %150 citations counted in INSPIRE as of 13 May 2017



\bibitem{Stave08} 
  S.~Stave {\it et al.} [A1 Collaboration],
  %``Measurements of the gamma* p ---> Delta Reaction At Low Q**2: Probing the Mesonic Contribution,''
  Phys.\ Rev.\ C {\bf 78}, 025209 (2008).
  %% doi:10.1103/PhysRevC.78.025209
  %%[arXiv:0803.2476 [hep-ex]].
  %%CITATION = doi:10.1103/PhysRevC.78.025209;%%
  %23 citations counted in INSPIRE as of 25 Dec 2015



\bibitem{Buchmann00b}
  A.~J.~Buchmann and E.~M.~Henley,
  %``Intrinsic quadrupole moment of the nucleon,''
  Phys.\ Rev.\ C {\bf 63}, 015202 (2000).
  %%[arXiv:hep-ph/0101027].
  %%CITATION = HEP-PH 0101027;%%



\bibitem{Tiator04} 
  L.~Tiator, D.~Drechsel, S.~Kamalov, M.~M.~Giannini, E.~Santopinto and A.~Vassallo,
  %``Electroproduction of nucleon resonances,''
  Eur.\ Phys.\ J.\ A {\bf 19}, 55 (2004).
  %% doi:10.1140/epjad/s2004-03-009-9
  %%[nucl-th/0310041].
  %%CITATION = doi:10.1140/epjad/s2004-03-009-9;%%
  %76 citations counted in INSPIRE as of 13 May 2017




\bibitem{SatoLee} 
  T.~Sato and T.~S.~H.~Lee,
  %``Dynamical study of the Delta excitation in N (e, e-prime pi) reactions,''
  Phys.\ Rev.\ C {\bf 63}, 055201 (2001).
  %%doi:10.1103/PhysRevC.63.055201
  %%[nucl-th/0010025].
  %%CITATION = doi:10.1103/PhysRevC.63.055201;%%
  %200 citations counted in INSPIRE as of 31 Aug 2017




\bibitem{QpionCloud-1}
  %\bibitem{Fiolhais96} 
  M.~Fiolhais, B.~Golli and S.~\v{S}irca,
  %``The Role of the pion cloud in electroproduction of the Delta (1232),''
  Phys.\ Lett.\ B {\bf 373}, 229 (1996).
  %% doi:10.1016/0370-2693(96)00130-X
  %% [hep-ph/9601379].
  %%CITATION = doi:10.1016/0370-2693(96)00130-X;%%
  %43 citations counted in INSPIRE as of 13 May 2016

\bibitem{QpionCloud-2}
  %\bibitem{QpionCloud2}
  %\bibitem{Lu97} 
  D.~H.~Lu, A.~W.~Thomas and A.~G.~Williams,
  %``A Chiral bag model approach to delta electroproduction,''
  Phys.\ Rev.\ C {\bf 55}, 3108 (1997).
  %% doi:10.1103/PhysRevC.55.3108
  %%[nucl-th/9612017].
  %%CITATION = doi:10.1103/PhysRevC.55.3108;%%
  %59 citations counted in INSPIRE as of 13 May 2016

\bibitem{QpionCloud-3}
  %%\bibitem{Dong01} 
  Y.~B.~Dong, K.~Shimizu and A.~Faessler,
  %``Meson cloud and electroproduction of the Delta(1232) resonance in a relativistic quark model approach,''
  Nucl.\ Phys.\ A {\bf 689}, 889 (2001).
  %%doi:10.1016/S0375-9474(00)00702-8
  %%CITATION = doi:10.1016/S0375-9474(00)00702-8;%%
  %5 citations counted in INSPIRE as of 15 Feb 2018






\bibitem{Kamalov01} 
  S.~S.~Kamalov, S.~N.~Yang, D.~Drechsel, O.~Hanstein and L.~Tiator,
  %``Gamma* N ---> Delta transition form-factors: A New analysis of the JLab data on p (e, e-prime p) pi0 at Q**2=(2.8-(GeV/c)**2 and 4.0-(GeV/c)**2),''
  Phys.\ Rev.\ C {\bf 64}, 032201 (2001).
  %% doi:10.1103/PhysRevC.64.032201
  %%[nucl-th/0006068].
  %%CITATION = doi:10.1103/PhysRevC.64.032201;%%
  %142 citations counted in INSPIRE as of 13 May 2017






\bibitem{Pascalutsa07a} 
  V.~Pascalutsa and M.~Vanderhaeghen,
  %``Large-N(c) relations for the electromagnetic N to Delta(1232) transition,''
  Phys.\ Rev.\ D {\bf 76}, 111501 (2007).
  %% doi:10.1103/PhysRevD.76.111501
  %%[arXiv:0711.0147 [hep-ph]].
  %%CITATION = doi:10.1103/PhysRevD.76.111501;%%
  %16 citations counted in INSPIRE as of 10 May 2016








\bibitem{Buchmann97a} 
  A.~J.~Buchmann, E.~Hernandez and A.~Faessler,
  %``Electromagnetic properties of the Delta (1232),''
  Phys.\ Rev.\ C {\bf 55}, 448 (1997).
  %% doi:10.1103/PhysRevC.55.448
  %% [nucl-th/9610040].
  %%CITATION = doi:10.1103/PhysRevC.55.448;%%
  %120 citations counted in INSPIRE as of 02 Apr 2016



\bibitem{Grabmayr01} 
  P.~Grabmayr and A.~J.~Buchmann,
  %``Moments of the neutron charge form-factor and the N ---> Delta quadrupole transition,''
  Phys.\ Rev.\ Lett.\  {\bf 86}, 2237 (2001).
  %% doi:10.1103/PhysRevLett.86.2237
  %%[hep-ph/0104203].
  %%CITATION = doi:10.1103/PhysRevLett.86.2237;%%
  %22 citations counted in INSPIRE as of 25 Mar 2016



\bibitem{Buchmann04} 
  A.~J.~Buchmann,
  %``Electromagnetic N ---> Delta transition and neutron form-factors,''
  Phys.\ Rev.\ Lett.\  {\bf 93}, 212301 (2004).
  %% doi:10.1103/PhysRevLett.93.212301
  %% [hep-ph/0412421].
  %%CITATION = doi:10.1103/PhysRevLett.93.212301;%%
  %34 citations counted in INSPIRE as of 25 Mar 2016






\bibitem{Buchmann02} 
  A.~J.~Buchmann, J.~A.~Hester and R.~F.~Lebed,
  %``Quadrupole moments of N and Delta in the 1 / N(c) expansion,''
  Phys.\ Rev.\ D {\bf 66}, 056002 (2002).
  %% doi:10.1103/PhysRevD.66.056002
  %% [hep-ph/0205108].
  %%CITATION = doi:10.1103/PhysRevD.66.056002;%%
  %47 citations counted in INSPIRE as of 01 Apr 2016
  %% LARGE NC







\bibitem{Buchmann09a} 
  A.~J.~Buchmann,
  %``Non-spherical proton shape and hydrogen hyperfine splitting,''
  Can.\ J.\ Phys.\  {\bf 87}, 773 (2009).
  %% doi:10.1139/P09-059
  %% [arXiv:0910.4747 [physics.atom-ph]].
  %%CITATION = doi:10.1139/P09-059;%%



\bibitem{Siegert-ND} 
  G.~Ramalho,
  %``Parametrizations of the $\gamma^\ast N \to \Delta(1232)$ quadrupole form factors and Siegert’s theorem,''
  Phys.\ Rev.\ D {\bf 94}, 114001 (2016).
  %% doi:10.1103/PhysRevD.94.114001
  %%[arXiv:1606.03042 [hep-ph]].
  %%CITATION = doi:10.1103/PhysRevD.94.114001;%%
  %4 citations counted in INSPIRE as of 13 May 2017






\bibitem{Blomberg16a}
  A.~Blomberg {\it et al.},
  %``Electroexcitation of the $\Delta^{+}(1232)$ at low momentum transfer,''
  Phys.\ Lett.\ B {\bf 760}, 267 (2016).
  %% doi:10.1016/j.physletb.2016.06.076
  %% [arXiv:1509.00780 [nucl-ex]].
  %%CITATION = doi:10.1016/j.physletb.2016.06.076;%%




\bibitem{Tiator-1}
  L.~Tiator and S.~Kamalov,
  %``Nucleon resonance excitation with virtual photons,''
  AIP Conf.\ Proc.\  {\bf 904}, 191 (2007).
  %% doi:10.1063/1.2734304
  %% [nucl-th/0610113];
  %%CITATION = doi:10.1063/1.2734304;%%
  %4 citations counted in INSPIRE as of 04 Aug 2017 

\bibitem{Tiator-2}
  L.~Tiator, D.~Drechsel, S.~S.~Kamalov and M.~Vanderhaeghen,
  %``Electromagnetic Excitation of Nucleon Resonances,''
  Eur.\ Phys.\ J.\ ST {\bf 198}, 141 (2011).
  %% doi:10.1140/epjst/e2011-01488-9
  %% [arXiv:1109.6745 [nucl-th]].
  %%CITATION = doi:10.1140/epjst/e2011-01488-9;%%
  %50 citations counted in INSPIRE as of 08 Nov 2017





\bibitem{Drechsel2007} 
  D.~Drechsel, S.~S.~Kamalov and L.~Tiator,
  %``Unitary Isobar Model - MAID2007,''
  Eur.\ Phys.\ J.\ A {\bf 34}, 69 (2007).
  %% doi:10.1140/epja/i2007-10490-6
  %% [arXiv:0710.0306 [nucl-th]].
  %%CITATION = doi:10.1140/epja/i2007-10490-6;%%
  %251 citations counted in INSPIRE as of 04 Dec 2015





\bibitem{SiegertD} 
  G.~Ramalho,
  %``Improved empirical parametrizations of the $\gamma^\ast N \to \Delta(1232)$ and $\gamma^\ast N \to N(1520)$ transition amplitudes and the Siegert's theorem,''
  Phys.\ Rev.\ D {\bf 93}, 113012 (2016).
  %% doi:10.1103/PhysRevD.93.113012
  %% [arXiv:1602.03832 [hep-ph]].
  %%CITATION = doi:10.1103/PhysRevD.93.113012;%%
  %3 citations counted in INSPIRE as of 21 Aug 2016



\bibitem{Siegert} 
  G.~Ramalho,
  %``Improved empirical parametrizations of the $\gamma^\ast N \to N(1535)$ transition amplitudes and the Siegert's theorem,''
  Phys.\ Lett.\ B {\bf 759}, 126 (2016).
  %% doi:10.1016/j.physletb.2016.05.060
  %% [arXiv:1602.03444 [hep-ph]].
  %%CITATION = doi:10.1016/j.physletb.2016.05.060;%%
  %2 citations counted in INSPIRE as of 01 Jun 2016









% REM, RSM data



\bibitem{Sparveris13} 
  N.~Sparveris {\it et al.},
  %``Measurements of the $\gamma$*p $\rightarrow$  $\Delta$ reaction at low Q$^{2}$,''
  Eur.\ Phys.\ J.\ A {\bf 49}, 136 (2013).
  %%doi:10.1140/epja/i2013-13136-2
  %% [arXiv:1307.0751 [nucl-ex]].
  %%CITATION = doi:10.1140/epja/i2013-13136-2;%%
  %2 citations counted in INSPIRE as of 03 Sep 2017


\bibitem{MIT_data}
  %\bibitem{Sparveris05} 
  N.~F.~Sparveris {\it et al.} [OOPS Collaboration],
  %``Investigation of the conjectured nucleon deformation at low momentum transfer,''
  Phys.\ Rev.\ Lett.\  {\bf 94}, 022003 (2005).
  %% doi:10.1103/PhysRevLett.94.022003
  %% [nucl-ex/0408003].








\bibitem{Jlab_data1} 
  %\bibitem{Kelly07} 
  J.~J.~Kelly {\it et al.},
  %``Recoil polarization measurements for neutral pion electroproduction at Q**2 = 1 (GeV/c)**2 near the Delta resonance,''
  Phys.\ Rev.\ C {\bf 75}, 025201 (2007).
  %% doi:10.1103/PhysRevC.75.025201
  %% [nucl-ex/0509004];
  %\bibitem{Aznauryan09}

\bibitem{Jlab_data2} 
  I.~G.~Aznauryan {\it et al.} [CLAS Collaboration],
  %``Electroexcitation of nucleon resonances from CLAS data on single pion electroproduction,''
  Phys.\ Rev.\ C {\bf 80}, 055203 (2009).
  %% doi:10.1103/PhysRevC.80.055203
  %%[arXiv:0909.2349 [nucl-ex]].
  %%CITATION = doi:10.1103/PhysRevC.80.055203;%%
  %122 citations counted in INSPIRE as of 25 Dec 2015
  % FROLOV Q2 > 2 GeV^2


\bibitem{PDG} 
  K.~A.~Olive {\it et al.} [Particle Data Group Collaboration],
  %``Review of Particle Physics,''
  Chin.\ Phys.\ C {\bf 38}, 090001 (2014).
  %% doi:10.1088/1674-1137/38/9/090001
  %%CITATION = doi:10.1088/1674-1137/38/9/090001;%%
  %2596 citations counted in INSPIRE as of 25 Dec 2015




\bibitem{IsgurRefs-1} 
  N.~Isgur and G.~Karl,
  %``Positive Parity Excited Baryons in a Quark Model with Hyperfine Interactions,''
  Phys.\ Rev.\ D {\bf 19}, 2653 (1979)
  Erratum: [Phys.\ Rev.\ D {\bf 23}, 817 (1981)].
  %% doi:10.1103/PhysRevD.23.817.2, 10.1103/PhysRevD.19.2653
  %%CITATION = doi:10.1103/PhysRevD.23.817.2, 10.1103/PhysRevD.19.2653;%%
  %972 citations counted in INSPIRE as of 13 May 2017
  %%

\bibitem{IsgurRefs-2} 
  N.~Isgur and G.~Karl,
  %``Ground State Baryons in a Quark Model with Hyperfine Interactions,''
  Phys.\ Rev.\ D {\bf 20}, 1191 (1979).
  %%doi:10.1103/PhysRevD.20.1191
  %%CITATION = doi:10.1103/PhysRevD.20.1191;%%
  %557 citations counted in INSPIRE as of 13 May 2017


\bibitem{IsgurRefs-3} 
  N.~Isgur, G.~Karl and D.~W.~L.~Sprung,
  %``The Neutron Charge Form-factor in a Quark Model With Hyperfine Interactions,''
  Phys.\ Rev.\ D {\bf 23}, 163 (1981).
  %% doi:10.1103/PhysRevD.23.163
  %%CITATION = doi:10.1103/PhysRevD.23.163;%%
  %98 citations counted in INSPIRE as of 13 May 2017








\bibitem{Dillon99} 
  G.~Dillon and G.~Morpurgo,
  %``General QCD parametrization of the charge radii of p, n and Delta+,''
  Phys.\ Lett.\ B {\bf 448}, 107 (1999).
  %% doi:10.1016/S0370-2693(99)00027-1
  %%CITATION = doi:10.1016/S0370-2693(99)00027-1;%%
  %32 citations counted in INSPIRE as of 12 Aug 2016



\bibitem{Buchmann02b} 
  A.~J.~Buchmann and E.~M.~Henley,
  %``Quadrupole moments of baryons,''
  Phys.\ Rev.\ D {\bf 65}, 073017 (2002).
  %% doi:10.1103/PhysRevD.65.073017
  %%CITATION = doi:10.1103/PhysRevD.65.073017;%%
  %43 citations counted in INSPIRE as of 12 Aug 2016



\bibitem{Jenkins02} 
  E.~E.~Jenkins, X.~Ji and A.~V.~Manohar,
  %``Delta ---> N gamma in large N(c) QCD,''
  Phys.\ Rev.\ Lett.\  {\bf 89}, 242001 (2002).
  %% doi:10.1103/PhysRevLett.89.242001
  %% [hep-ph/0207092].
  %%CITATION = doi:10.1103/PhysRevLett.89.242001;%%
  %52 citations counted in INSPIRE as of 10 May 2016



%\vspace{.5cm}






%  QM vs Siegert's theorem

% \bibitem{Siegert-QM}

\bibitem{Drechsel84} 
   D.~Drechsel and M.~M.~Giannini,
  %``A Note On The Quadrupole (n ---> Delta) Transition Amplitude In Quark Models,''
  Phys.\ Lett.\  {\bf 143B}, 329 (1984).
  %% doi:10.1016/0370-2693(84)91476-X
  %%CITATION = doi:10.1016/0370-2693(84)91476-X;%%
  %57 citations counted in INSPIRE as of 13 May 2017

\bibitem{Weyrauch86} 
   M.~Weyrauch and H.~J.~Weber,
  %``Color Hyperfine Versus Pion Dynamics in the $N(939) - \Delta(1232)$ Baryons,''
  Phys.\ Lett.\ B {\bf 171}, 13 (1986)
  [Phys.\ Lett.\ B {\bf 181}, 415 (1986)].
  %% doi:10.1016/0370-2693(86)90988-3
  %%CITATION = doi:10.1016/0370-2693(86)90988-3;%%
  %30 citations counted in INSPIRE as of 01 Jan 2016

\bibitem{Bourdeau87} 
  M.~Bourdeau and N.~C.~Mukhopadhyay,
  %``Color Magnetism and the Helicity Zero ($\gamma$ (Neutrino) $N \to \Delta$) Transition Amplitude,''
  Phys.\ Rev.\ Lett.\  {\bf 58}, 976 (1987).
  %% doi:10.1103/PhysRevLett.58.976
  %%CITATION = doi:10.1103/PhysRevLett.58.976;%%
  %39 citations counted in INSPIRE as of 01 Jan 2016



\bibitem{Buchmann98} 
  A.~J.~Buchmann, U.~Meyer, A.~Faessler and E.~Hernandez,
  %``N --> Delta (1232) E2 transition and Siegert's theorem,''
  Phys.\ Rev.\ C {\bf 58}, 2478 (1998).
  %% doi:10.1103/PhysRevC.58.2478
  %%CITATION = doi:10.1103/PhysRevC.58.2478;%%
  %35 citations counted in INSPIRE as of 23 Sep 2017





% Galster parametrization

\bibitem{Galster71} 
  S.~Galster, H.~Klein, J.~Moritz, K.~H.~Schmidt, D.~Wegener and J.~Bleckwenn,
  %``Elastic electron-deuteron scattering and the electric neutron form factor at four-momentum transfers 5fm$^{-2} < q^2 < 14$fm$^{-2}$,''
  Nucl.\ Phys.\ B {\bf 32}, 221 (1971).
  %% doi:10.1016/0550-3213(71)90068-X
  %%CITATION = doi:10.1016/0550-3213(71)90068-X;%%
  %589 citations counted in INSPIRE as of 24 May 2017






\bibitem{Kelly02} 
  J.~J.~Kelly,
  %``Nucleon charge and magnetization densities from Sachs form-factors,''
  Phys.\ Rev.\ C {\bf 66}, 065203 (2002).
  %% doi:10.1103/PhysRevC.66.065203
  %% [hep-ph/0204239].
  %%CITATION = doi:10.1103/PhysRevC.66.065203;%%
  %112 citations counted in INSPIRE as of 24 May 2017
  % LONG PAPER


\bibitem{Friedrich03} 
  J.~Friedrich and T.~Walcher,
  %``A Coherent interpretation of the form-factors of the nucleon in terms of a pion cloud and constituent quarks,''
  Eur.\ Phys.\ J.\ A {\bf 17}, 607 (2003).
  %%doi:10.1140/epja/i2003-10025-3
  %% [hep-ph/0303054].
  %%CITATION = doi:10.1140/epja/i2003-10025-3;%%
  %182 citations counted in INSPIRE as of 24 May 2017






\bibitem{Kaskulov04} 
  M.~M.~Kaskulov and P.~Grabmayr,
  %``The Electric form-factor of the neutron and its chiral content,''
  Eur.\ Phys.\ J.\ A {\bf 19}, 157 (2004).
  %% doi:10.1140/epja/i2003-10141-0
  %%[nucl-th/0308015].
  %%CITATION = doi:10.1140/epja/i2003-10141-0;%%
  %7 citations counted in INSPIRE as of 03 Oct 2017




\bibitem{Bertozzi72} 
  W.~Bertozzi, J.~Friar, J.~Heisenberg and J.~W.~Negele,
  %``Contributions of neutrons to elastic electron scattering from nuclei,''
  Phys.\ Lett.\  {\bf 41B}, 408 (1972).
  %% doi:10.1016/0370-2693(72)90662-4
  %%CITATION = doi:10.1016/0370-2693(72)90662-4;%%
  %161 citations counted in INSPIRE as of 24 May 2017


\bibitem{Platchkov90} 
  S.~Platchkov {\it et al.},
  %``Deuteron A(q**2) Structure Function And The Neutron Electric Form-factor,''
  Nucl.\ Phys.\ A {\bf 510}, 740 (1990).
  %% doi:10.1016/0375-9474(90)90358-S
  %%CITATION = doi:10.1016/0375-9474(90)90358-S;%%
  %308 citations counted in INSPIRE as of 24 May 2017





\bibitem{Gentile11} 
  T.~R.~Gentile and C.~B.~Crawford,
  %``Neutron charge radius and the neutron electric form factor,''
  Phys.\ Rev.\ C {\bf 83}, 055203 (2011).
  %% doi:10.1103/PhysRevC.83.055203
  %%CITATION = doi:10.1103/PhysRevC.83.055203;%%
  %11 citations counted in INSPIRE as of 24 May 2017









\bibitem{Hammer04} 
  H.~W.~Hammer and U.~G.~Meissner,
  %``Updated dispersion theoretical analysis of the nucleon electromagnetic form-factors,''
  Eur.\ Phys.\ J.\ A {\bf 20}, no. 3, 469 (2004).
  %%doi:10.1140/epja/i2003-10223-y
  %% [hep-ph/0312081].
  %%CITATION = doi:10.1140/epja/i2003-10223-y;%%
  %70 citations counted in INSPIRE as of 14 Feb 2018



\bibitem{Belushkin07} 
  M.~A.~Belushkin, H.-W.~Hammer and U.-G.~Meissner,
  %``Dispersion analysis of the nucleon form-factors including meson continua,''
  Phys.\ Rev.\ C {\bf 75}, 035202 (2007).
  %doi:10.1103/PhysRevC.75.035202
  %% [hep-ph/0608337].
  %%CITATION = doi:10.1103/PhysRevC.75.035202;%%
  %160 citations counted in INSPIRE as of 14 Feb 2018



\bibitem{Lorenz12} 
  I.~T.~Lorenz, H.-W.~Hammer and U.~G.~Meissner,
  %``The size of the proton - closing in on the radius puzzle,''
  Eur.\ Phys.\ J.\ A {\bf 48}, 151 (2012).
  %%doi:10.1140/epja/i2012-12151-1
  %% [arXiv:1205.6628 [hep-ph]].
  %%CITATION = doi:10.1140/epja/i2012-12151-1;%%
  %76 citations counted in INSPIRE as of 14 Feb 2018












\bibitem{InPreparation} 
  %G.~Ramalho, work in preparation.
   G.~Ramalho,
  %``Combined parametrization of the neutron electric form factor and the $\gamma^\ast N \to \Delta(1232)$ quadrupole form factors,''
  arXiv:1710.10527 [hep-ph].
  %%CITATION = ARXIV:1710.10527;%%








\bibitem{Lichtenberg78} 
  D.~B.~Lichtenberg,
  {\it Unitary Symmetry and Elementary Particles},
  %``Unitary Symmetry and Elementary Particles,''
  Academic Press/New York 1978. %257p
  %4 citations counted in INSPIRE as of 08 Nov 2017


\bibitem{Close79} 
  F.~E.~Close,
  {\it An Introduction to Quarks and Partons},
  %``An Introduction to Quarks and Partons,''
  Academic Press/London 1979. %481p
  %78 citations counted in INSPIRE as of 08 Nov 2017


\bibitem{Buchmann91a} 
  A.~Buchmann, E.~Hernandez and K.~Yazaki,
  %``Gluon and pion exchange currents in the nucleon,''
  Phys.\ Lett.\ B {\bf 269}, 35 (1991).
  %% doi:10.1016/0370-2693(91)91448-5
  %%CITATION = doi:10.1016/0370-2693(91)91448-5;%%
  %89 citations counted in INSPIRE as of 09 Nov 2017


\bibitem{Christov96} 
  C.~V.~Christov, A.~Blotz, H.~C.~Kim, P.~Pobylitsa, T.~Watabe, T.~Meissner, 
  E.~Ruiz Arriola and K.~Goeke,
  %``Baryons as nontopological chiral solitons,''
  Prog.\ Part.\ Nucl.\ Phys.\  {\bf 37}, 91 (1996).
  %%doi:10.1016/0146-6410(96)00057-9
  %% [hep-ph/9604441].
  %%CITATION = doi:10.1016/0146-6410(96)00057-9;%%
  %277 citations counted in INSPIRE as of 09 Nov 2017




\bibitem{Lu98} 
  D.~H.~Lu, A.~W.~Thomas and A.~G.~Williams,
  %``Electromagnetic form-factors of the nucleon in an improved quark model,''
  Phys.\ Rev.\ C {\bf 57}, 2628 (1998).
  %%doi:10.1103/PhysRevC.57.2628
  %% [nucl-th/9706019].
  %%CITATION = doi:10.1103/PhysRevC.57.2628;%%
  %103 citations counted in INSPIRE as of 09 Nov 2017




\newpage

\bibitem{Nucleon} 
  F.~Gross, G.~Ramalho and M.~T.~Pe\~na,
  %``A Pure S-wave covariant model for the nucleon,''
  Phys.\ Rev.\ C {\bf 77}, 015202 (2008).
  %% doi:10.1103/PhysRevC.77.015202
  %% [nucl-th/0606029].
  %%CITATION = doi:10.1103/PhysRevC.77.015202;%%
  %66 citations counted in INSPIRE as of 12 May 2016




\bibitem{Omega} 
  G.~Ramalho, K.~Tsushima and F.~Gross,
  %``A Relativistic quark model for the Omega- electromagnetic form factors,''
  Phys.\ Rev.\ D {\bf 80}, 033004 (2009).
  %% doi:10.1103/PhysRevD.80.033004
  %% [arXiv:0907.1060 [hep-ph]].
  %%CITATION = doi:10.1103/PhysRevD.80.033004;%%
  %36 citations counted in INSPIRE as of 21 Aug 2016





%\newpage


\bibitem{Alexandrou08} 
  C.~Alexandrou, G.~Koutsou, H.~Neff, J.~W.~Negele, W.~Schroers and A.~Tsapalis,
  %``Nucleon to delta electromagnetic transition form factors in lattice QCD,''
  Phys.\ Rev.\ D {\bf 77}, 085012 (2008).
  %% doi:10.1103/PhysRevD.77.085012
  %%[arXiv:0710.4621 [hep-lat]].
  %%CITATION = doi:10.1103/PhysRevD.77.085012;%%
  %60 citations counted in INSPIRE as of 10 May 2016


%\newpage



\bibitem{Carlson-1} 
  C.~E.~Carlson and N.~C.~Mukhopadhyay,
  %``Approach to perturbative results in the N - Delta transition,''
  Phys.\ Rev.\ Lett.\  {\bf 81}, 2646 (1998).
  %% doi:10.1103/PhysRevLett.81.2646
  %%[hep-ph/9804356];


\bibitem{Carlson-2} 
   C.~E.~Carlson,
  %``Electromagnetic N - Delta Transition At High Q**2,''
  Phys.\ Rev.\ D {\bf 34}, 2704 (1986).
  %% doi:10.1103/PhysRevD.34.2704
  %%CITATION = doi:10.1103/PhysRevD.34.2704;%%
  %159 citations counted in INSPIRE as of 13 Nov 2016




\bibitem{Alexandrou11} 
  C.~Alexandrou, G.~Koutsou, J.~W.~Negele, Y.~Proestos and A.~Tsapalis,
  %``Nucleon to Delta transition form factors with $N_F=2+1$ domain wall fermions,''
  Phys.\ Rev.\ D {\bf 83}, 014501 (2011).
  %% doi:10.1103/PhysRevD.83.014501
  %% [arXiv:1011.3233 [hep-lat]].
  %%CITATION = doi:10.1103/PhysRevD.83.014501;%%
  %33 citations counted in INSPIRE as of 06 Sep 2017

\bibitem{Pascalutsa06} 
  V.~Pascalutsa and M.~Vanderhaeghen,
  %``The Nucleon and delta-resonance masses in relativistic chiral effective-field theory,''
  Phys.\ Lett.\ B {\bf 636}, 31 (2006).
  %% doi:10.1016/j.physletb.2006.03.023
  %% [hep-ph/0511261].
  %%CITATION = doi:10.1016/j.physletb.2006.03.023;%%
  %50 citations counted in INSPIRE as of 06 Sep 2017

\bibitem{Gail06} 
  T.~A.~Gail and T.~R.~Hemmert,
  %``Signatures of chiral dynamics in the nucleon to delta transition,''
  Eur.\ Phys.\ J.\ A {\bf 28}, 91 (2006).
  %%doi:10.1140/epja/i2006-10023-y
  %%[nucl-th/0512082].
  %%CITATION = doi:10.1140/epja/i2006-10023-y;%%
  %45 citations counted in INSPIRE as of 06 Sep 2017





\bibitem{Agadjanov14} 
  A.~Agadjanov, V.~Bernard, U.~G.~Meißner and A.~Rusetsky,
  %``A framework for the calculation of the $ΔNγ^⁎$ transition form factors on the lattice,''
  Nucl.\ Phys.\ B {\bf 886}, 1199 (2014).
  %%doi:10.1016/j.nuclphysb.2014.07.023
  %% [arXiv:1405.3476 [hep-lat]].
  %%CITATION = doi:10.1016/j.nuclphysb.2014.07.023;%%
  %37 citations counted in INSPIRE as of 14 Feb 2018

\bibitem{Faessler06} 
  A.~Faessler, T.~Gutsche, B.~R.~Holstein, V.~E.~Lyubovitskij, D.~Nicmorus and K.~Pumsa-ard,
  %``Light baryon magnetic moments and N ---> Delta gamma transition in a Lorentz covariant chiral quark approach,''
  Phys.\ Rev.\ D {\bf 74}, 074010 (2006).
  %%doi:10.1103/PhysRevD.74.074010
  %% [hep-ph/0608015].
  %%CITATION = doi:10.1103/PhysRevD.74.074010;%%
  %62 citations counted in INSPIRE as of 14 Feb 2018




\bibitem{firstMeas-1}
     R.~Haidan, PhD Thesis, DESY report No.~F21-79-03 (1979).
     %


\bibitem{firstMeas-2}
    F.~Kalleicher, U.~Dittmayer, R.~W.~Gothe, H.~Putsch, T.~Reichelt, B.~Schoch and M.~Wilhelm,
  %``The determination of sigma(LT)/sigma(TT) in electropion production in the Delta resonance region,''
  Z.\ Phys.\ A {\bf 359}, 201 (1997).
  %doi:10.1007/s002180050387
  %%CITATION = doi:10.1007/s002180050387;%%
  %52 citations counted in INSPIRE as of 30 Aug 2017


\bibitem{firstMeas-3}
  %% \bibitem{Frolov99} 
  V.~V.~Frolov {\it et al.},
  %``Electroproduction of the Delta (1232) resonance at high momentum transfer,''
  Phys.\ Rev.\ Lett.\  {\bf 82}, 45 (1999).
  %% doi:10.1103/PhysRevLett.82.45
  %%[hep-ex/9808024].
  %%CITATION = doi:10.1103/PhysRevLett.82.45;%%
  %202 citations counted in INSPIRE as of 27 Aug 2017



\end{thebibliography}
\end{document}